# Casimir effect of an ideal Bose gas trapped in a generic power-law potential

Tongling Lin[1,2,3], Guozhen Su[1], Qiuping A. Wang[2,3] and Jincan Chen[1 (a)]

[1] *Department of Physics, Xiamen University, Xiamen 361005, People's Republic of China*
[2] *LUNAM Université, ISMANS, Laboratoire de Physique Statistique et Systèmes Complexes, 44, Avenue F.A. Bartholdi, 72000 Le Mans, France*
[3] *LPEC, Université du Maine, Ave. O. Messiaen, 72035 Le Mans, France*

**Abstract** – The Casimir effect of an ideal Bose gas trapped in a generic power-law potential and confined between two slabs with Dirichlet, Neumann, and periodic boundary conditions is investigated systematically, based on the grand potential of the ideal Bose gas, the Casimir potential and force are calculated. The scaling function is obtained and discussed. The special cases of free and harmonic potentials are also discussed. It is found that when $T \leq T_c$ (where $T_c$ is the critical temperature of Bose-Einstein condensation), the Casimir force is a power-law decay function; when $T > T_c$, the Casimir force is an exponential decay function; and when $T \gg T_c$, the Casimir force vanishes.

The original Casimir effect [1] shows that zero-temperature quantum fluctuations in an electromagnetic vacuum give rise to an attractive force between two closely spaced perfectly conducting plates. It is a pure quantum effect because there is no force between the plates in classical electrodynamics. The Casimir effect for massive quantum particles is less explored than its counterpart for the photon gas [2], but the Casimir effect in a confined Bose-Einstein condensate (BEC) system caused by the quantum fluctuations of the ground state at zero temperature or thermal fluctuations at finite temperature has recently attracted considerable interest [2-15]. The Casimir effect caused by thermal fluctuations in a Bose gas confined by two slabs was studied under the Dirichlet, Neumann, and periodic boundary conditions by Martin and Zagrebnov [2]. The asymptotic expressions of the grand potential with a universal Casimir term [2], the relationship between the thermodynamic Casimir effect in the Bose gas slabs and the critical Casimir forces [3-5], and the scaling function [6] for an ideal Bose gas in the case of the Dirichlet boundary condition were obtained. The Casimir effect of an ideal Bose gas was also considered at finite temperatures with or without traps for the Dirichlet boundary condition [8, 9].

By using the field-theoretical method [6, 7] for the plate geometry instead of the thermodynamic method, it became possible to consider the Casimir effect of a weakly interacting Bose gas. Recently, the Casimir force due to zero-temperature quantum fluctuations and thermal fluctuations at finite temperature of a weakly-interacting dilute BEC confined by a pair of parallel plates with Dirichlet and periodic boundary conditions was also investigated [10-13]. In addition, it has been noted [10, 12] that the quasiparticle vacuum in a zero-temperature dilute weakly-interacting BEC should give rise to a measurable Casimir force.

The purpose of this article is to investigate the behaviour of the Casimir effect in an ideal Bose gas confined by two slabs in the presence of a generic power-law potential trap [16, 17]. Starting from the grand potential of the system, we give a significant description for the Casimir potential in the Dirichlet, Neumann, and periodic boundary conditions. The scaling functions and Casimir forces are independently obtained in the three types of boundary conditions. The special cases of free and harmonic potentials are also discussed.

Let us consider an ideal Bose gas confined between two infinite parallel slabs that are separated by a distance $d$ in the $z$ direction. The Bose gas is trapped in a generic power-law potential in the x-y plane. The Hamiltonian of the Bose gas in the x-y plane can be expressed as follows:

$$H = \frac{p^2}{2m} + \sum_{k=1}^{2} U_k \left| \frac{r_k}{L_k} \right|^{t_k}, \qquad (1)$$

where $U_k$, $L_k$ and $t_k$ are all positive constants, $m$ is the mass of a particle, $p = (\Sigma_{k=1}^{2} p_k^2)^{1/2}$, and $p_k$ and $r_k$ are the $k$th component of the momentum and the $k$th coordinate in the x-y plane, respectively.

It is assumed that the particle number between the two slabs is large, and the level spacing of the transverse motion (the motion of particles in the x-y plane), which is dependent on the parameters $U_k$, $L_k$, and $t_k$ and the mass $m$, is much smaller than the corresponding mean particle energy. In this case, the energy of the transverse motion can be treated as

---

[(a)]E-mail: jcchen@xmu.edu.cn

continuous and hence the sum over the possible levels can be well replaced by the integral with respect to the energy. In contrast, the level spacing of the longitudinal motion (the motion of particles in the $z$ direction) may be large for a distance $d$, so that the longitudinal motion should be described by quantum mechanics.

Based on the above assumption, the grand potential of the ideal Bose gas between two slabs is given by

$$\varphi_d(T,\mu) = \beta^{-1} \sum_{n=1}^{\infty} \int_0^{\infty} \ln(1 - z e^{-\beta \frac{\pi^2 \hbar^2}{2md^2} n^2} e^{-\beta \varepsilon}) D(\varepsilon) d\varepsilon \quad (2)$$

for the Dirichlet boundary condition in two slabs, where $\beta = 1/(k_B T)$, $k_B$ is Boltzmann's constant, $T$ is the temperature, $\hbar = h/2\pi$, $h$ is Planck's constant, $z = \exp(\beta \mu)$ is the fugacity, $\mu$ is the chemical potential,

$$D(\varepsilon) = \frac{\alpha}{\Gamma(\sigma)} \varepsilon^{\sigma-1} \quad (3)$$

is the density of states for the motion in the x-y plane [16, 17], $\alpha = \frac{2m}{\hbar^2 \pi} \prod_{k=1}^{2} \frac{L_k \Gamma(1/t_k + 1)}{U_k^{1/t_k}}$, $\Gamma(\sigma) = \int_0^{\infty} x^{\sigma-1} e^{-x} dx$ is the Gamma function, and $\sigma = 1 + \sum_{k=1}^{2} \frac{1}{t_k}$. Equation (2) does not include the ground-state term because it can be proven that the contribution of the ground state to the grand potential can be neglected at finite temperature. Substituting Eq. (3) into Eq. (2), the grand potential can be derived as follows:

$$\varphi_d(T,\mu) = -\frac{2m}{\hbar^2 \pi} \prod_{k=1}^{2} \frac{L_k \Gamma(1/t_k + 1)}{U_k^{1/t_k}} \beta^{-\sigma-1} \sum_{r=1}^{\infty} \frac{e^{\beta \mu r}}{r^{\sigma+1}} \sum_{n=1}^{\infty} e^{-\pi n^2 (r\pi(\lambda/d)^2/2)}, \quad (4)$$

where $\lambda = \hbar\sqrt{\beta/m}$ is the thermal wavelength. By using the Jacobi identity [18]

$$\sum_{n=1}^{\infty} e^{-\pi n^2 a} = (\frac{1}{2\sqrt{a}} - \frac{1}{2}) + \frac{1}{\sqrt{a}} \sum_{n=1}^{\infty} e^{-\pi n^2/a}, \quad a > 0, \quad (5)$$

$\varphi_d(T,\mu)$ can be expressed as follows:

$$\varphi_d(T,\mu) = -\frac{2m}{\hbar^2 \pi} \prod_{k=1}^{2} \frac{L_k \Gamma(1/t_k + 1)}{U_k^{1/t_k}} \beta^{-\sigma-1} \sum_{r=1}^{\infty} \frac{e^{\beta \mu r}}{r^{\sigma+1}} \quad (6)$$

$$\times \left[ \frac{d}{\lambda\sqrt{2\pi} r^{1/2}} - \frac{1}{2} + \frac{\sqrt{2} d}{\lambda\sqrt{\pi} r^{1/2}} \sum_{n=1}^{\infty} e^{-2(nd/\lambda)^2/r} \right]$$

It can be shown that the first and second terms in Eq. (6), which are similar to the bulk term $\varphi_{bulk}(T,\mu)$ and the surface term $\varphi_{surf}(T,\mu)$, respectively, of the grand potential described in Refs. [2, 8], do not contribute to the Casimir force, because the force due to the bulk term $\varphi_{bulk}(T,\mu)$ is counterbalanced by the same contribution acting from outside the slabs when they are immersed in the critical medium [4] and the surface term $\varphi_{surf}(T,\mu)$ does not change with the change of $d$. The last term in Eq. (6)

$$\varphi_{Cas}(T,\mu) = -(\frac{2}{\pi})^{3/2} \frac{m}{\hbar^2} \prod_{k=1}^{2} \frac{L_k \Gamma(1/t_k + 1)}{U_k^{1/t_k}} \beta^{-\sigma-1} \frac{d}{\lambda} \quad (7)$$

$$\times \sum_{r=1}^{\infty} \frac{e^{\beta \mu r}}{r^{\sigma+3/2}} \sum_{n=1}^{\infty} e^{-2(nd/\lambda)^2/r}$$

is just the Casimir potential, which in fact gives rise to a measurable effect [4]: a small displacement, $\delta d$, of one of the two slabs produces a change of $\delta\varphi_d$ in the grand potential and therefore results in a force $-\delta\varphi_d/\delta d$ acting on the slabs.

It is well known that BEC can occur in a Bose gas when $T \le T_C$. Thus, we discuss the Casimir effect in the cases of $T \le T_C$ ($\mu=0$) and $T > T_C$ ($\mu<0$).

(1) When $T \le T_C$, Eq. (7) can be simplified as follows:

$$\varphi_{Cas}(T,0) = C\beta^{-1} d^{-2\sigma} \sum_{n=1}^{\infty} (\lambda/d)^2 \sum_{r=1}^{\infty} \Phi_n((\lambda/d)^2 r), \quad (8)$$

where $\Phi_n(x) = x^{-\sigma-3/2} e^{-B_n/x}$, $B_n = 2n^2$, and $C = -(2/\pi)^{3/2} \frac{\hbar^{2\sigma-2}}{m^{\sigma-1}} \prod_{k=1}^{2} \frac{L_k \Gamma(1/t_k+1)}{U_k^{1/t_k}}$. We consider the case of a large $d$ asymptote to set $\lambda/d \ll 1$. In this case, the sum $\sum_{r=1}^{\infty}$ of the right-hand side of Eq. (8) can be converted into an integral; consequently, Eq. (8) can be expressed as follows:

$$\varphi_{Cas}(T,0) \quad (9)$$
$$= 2^{-\sigma-1/2} C\beta^{-1} \Gamma(\sigma+1/2) \zeta(2\sigma+1) d^{-2\sigma},$$
$$\propto 1/d^{2\sigma}$$

where $\zeta(\sigma) = \sum_{n=1}^{\infty} n^{-\sigma}$ is the Riemann zeta-function. From Eq. (9), the Casimir force $F_c(d,T,0) = -\partial\varphi_{Cas}(T,0)/\partial d \propto 1/d^{2\sigma+1}$ can be calculated. It can be found that the Casimir potential and force are the power-law functions of $d$, and they decay with different powers for different potentials.

(2) When $T > T_C$, the double sums in Eq. (7) for large $d$ are estimated as follows [2]:

$$\sum_{r=1}^{\infty} \frac{e^{\beta \mu r}}{r^{\sigma+3/2}} \sum_{n=1}^{\infty} e^{-2(nd/\lambda)^2/r} \quad (10)$$

$$\le \frac{\zeta(\sigma+3/2)}{e^{\sqrt{-8\beta\mu} d/\lambda} - 1} = O(e^{-\sqrt{-8\beta\mu} d/\lambda}),$$

Equation (10) implies that the Casimir potential $\varphi_{Cas}(T,\mu)$ and the Casimir force $F_c(d,T,\mu) = -\partial\varphi_{Cas}(T,\mu)/\partial d$ decay exponentially with $d$ when $T > T_C$, and they vanish when $T \gg T_C$.

The results obtained above are general, and they may be directly used to discuss the Casimir effect in special cases.

**a) The free Bose gas**

Letting $t_k = \infty$, which corresponds to the case of the absence of the external potential, we can obtain $\sigma = 1 + \sum_{k=1}^{2} \frac{1}{t_k} = 1$ and $C = -(2/\pi)^{3/2} L_1 L_2$. According to Eq. (7),

$$\beta d^2 \varphi_{Cas}(T,\mu) \qquad (11)$$

$$= C\left(\frac{d}{\lambda}\right)^3 \sum_{n,r=1}^{\infty} \frac{e^{\beta \mu r}}{r^{5/2}} e^{-2(nd/\lambda)^2/r}$$

$$= 2C \sum_{n=1}^{\infty} \int_0^{\infty} x^{-4} e^{-B_u x^2 - B_n/x^2} dx ,$$

$$= C \sum_{n=1}^{\infty} \sqrt{\frac{\pi}{B_u}} \frac{d^2}{dB_n^2} e^{-2\sqrt{B_u B_n}}$$

$$= -\frac{A}{8\pi} \sum_{n=1}^{\infty} \left(\frac{1+2un}{n^3}\right) e^{-2un} \equiv A\Theta_F(u)$$

where $u = (-2\beta\mu)^{1/2} d/\lambda \sim \frac{d}{\xi}$ [3], $\xi$ is the correlation length [6], $B_u = u^2/2$, $A = 4L_1 L_2$, and $\Theta_F(u)$ is the scaling function [3, 6] in the case of the free Bose gas, which is a continuous, negative, and monotonically increasing function of $u$. Thus, the Casimir force of the free ideal Bose gas can be obtained as follows:

$$F_c(d,T,\mu) = k_B T A [2\Theta_F(u) - u\Theta'_F(u)]/d^3 , \qquad (12)$$

where $\Theta'_F(u) = \frac{1}{2\pi} \sum_{n=1}^{\infty} (u/n) \exp(-2un)$.

When $T \leq T_c$, the chemical potential $\mu = 0$; consequently, $u = 0$, $\Theta_F(0) = -\zeta(3)/(8\pi) \equiv \Delta_{O,O}^{(1)}$, $\Theta'_F(0) = 0$, the correlated fluctuations become long-ranged [3], and the Casimir force per unit area is given by

$$f_c = \frac{F_c}{A} = -\frac{\zeta(3)}{4\pi} \frac{k_B T}{d^3} , \qquad (13)$$

where $\Delta_{O,O}^{(1)}$ is the Casimir amplitude given by Eq. (5.6) in Ref. [6] for $D=3$ and $N=2$. It can be seen from Eq. (13) that the Casimir force for the free ideal Bose gas varies with $d^{-3}$ and has a purely classical expression that is independent of $\hbar$ and $m$. It is seen from Eq. (13) that the Casimir force is proportional to an energy scale $k_B T$ when $T \leq T_c$. Thus, the Casimir force vanishes when $T \to 0$.

When $T > T_c$, the chemical potential $\mu \neq 0$, so that $u \gg 1$ and $\Theta_F(u \gg 1) \sim e^{-2u}$ for $d/\lambda \gg 1$. Therefore, $|\varphi_{Cas}(T,\mu)| \sim O(e^{-\sqrt{-8\beta\mu} d/\lambda})$ is expected to be exponentially small, and the Casimir force decays exponentially with $d$. When $T \gg T_c$, $\Theta_F(u \to \infty) = 0$. The correlated fluctuations between the two slabs are no longer long-ranged; therefore, the confining boundaries are subject to a vanishing Casimir force [3].

**b) The harmonically trapped Bose gas**

Letting $t_k = 2$ and $U_k/L_k^{t_k} = m\omega^2/2$, which corresponds to the case of the isotropic harmonic potential of angular frequency $\omega$, we can obtain $\sigma = 1 + \sum_{k=1}^{2} \frac{1}{t_k} = 2$ and $C = -\sqrt{\frac{2}{\pi}} \frac{\hbar^2}{m^2 \omega^2}$. In this case, the scaling function can be expressed as follows:

$$\beta d^4 \varphi_{Cas}(T,\mu) \qquad (14)$$

$$= C\left(\frac{d}{\lambda}\right)^5 \sum_{n,r=1}^{\infty} \frac{e^{\beta\mu r}}{r^{7/2}} e^{-2(nd/\lambda)^2/r}$$

$$= 2C \sum_{n=1}^{\infty} \int_0^{\infty} x^{-6} e^{-B_u x^2 - B_n/x^2} dx$$

$$= -C \sum_{n=1}^{\infty} \sqrt{\frac{\pi}{B_u}} \frac{d^3}{dB_n^3} e^{-2\sqrt{B_u B_n}}$$

$$= -\frac{3}{16} \frac{\hbar^2}{m^2\omega^2} \sum_{n=1}^{\infty} \left(\frac{1+2un+4/3 u^2 n^2}{n^5}\right) e^{-2un} \equiv \frac{\hbar^2}{m^2\omega^2} \Theta_H(u)$$

and the following Casimir force can be obtained:

$$F_c(d,T,\mu) = k_B T \frac{\hbar^2}{m^2\omega^2} [4\Theta_H(u) - u\Theta'_H(u)]/d^5 , \qquad (15)$$

where $\Theta_H(u)$ is the scaling function in the case of the harmonically trapped Bose gas, and $\Theta'_H(u) = \frac{1}{4} \sum_{n=1}^{\infty} [u(1+2un)/n^3] \exp(-2un)$.

When $T \leq T_c$, the chemical potential $\mu = 0$, so that $u = 0$, $\Theta_H(0) = -\frac{3}{16}\zeta(5)$, $\Theta'_H(0) = 0$, and the Casimir force is now given by the following:

$$F_c = -\frac{3}{4} \frac{\zeta(5)\hbar^2}{m^2\omega^2} \frac{k_B T}{d^5} . \qquad (16)$$

The Casimir force is inversely proportional to $d^5$ in the case of the harmonic potential, which is notably different from the case of the free Bose gas.

When $T > T_c$, the Casimir force for the harmonically trapped Bose gas decays exponentially, and it vanishes at the high temperature limit.

In addition to the Casimir effect in the case of the Dirichlet boundary condition discussed above, we can also discuss the Casimir effect in other boundary conditions in a straightforward fashion.

For the Neumann boundary condition, the sums in Eqs. (4) and (5) start from $n=0$, which implies only that the sign of the surface term changes. Therefore, the Casimir potential and force are the same as those in the case of the Dirichlet boundary condition.

For the periodic boundary condition, the Casimir potential can be derived as follows:

$$\varphi_{Cas}(T,\mu) \tag{17}$$
$$= -\left(\frac{2}{\pi}\right)^{3/2} \frac{m}{\hbar^2} \prod_{k=1}^{2} \frac{L_k \Gamma(1/t_k+1)}{U_k^{1/t_k}} \beta^{-\sigma-1} \frac{d}{\lambda}$$
$$\times \sum_{r=1}^{\infty} \frac{e^{\beta\mu r}}{r^{\sigma+3/2}} \sum_{n=1}^{\infty} e^{-2(\frac{nd}{2\lambda})^2/r}$$

and the scaling functions of the Casimir potential are given by

$$\beta d^2 \varphi_{Cas}(T,\mu) = -\frac{A}{\pi} \sum_{n=1}^{\infty} \left(\frac{1+un}{n^3}\right) e^{-un} \equiv A\Theta_{per,F}(u) \tag{18}$$

and

$$\beta d^4 \varphi_{Cas}(T,\mu) \tag{19}$$
$$= -\frac{6\hbar^2}{m^2\omega^2} \sum_{n=1}^{\infty} \left(\frac{1+un+u^2n^2/3}{n^5}\right) e^{-un} \equiv \frac{\hbar^2}{m^2\omega^2}\Theta_{per,H}(u)$$

for the free and harmonically trapped ideal Bose gas, respectively. It is easy to verify that when $u=0$, $\Theta_{per,F}(0) = -\zeta(3)/\pi = \Delta_{per}^{(1)}$, where $\Delta_{per}^{(1)}$ is the Casimir amplitude given by Eq. (5.7) in Ref. [6] for $D=3$ and $N=2$. From Eqs. (18) and (19), we can calculate that when $T \le T_c$, the Casimir forces per unit area in the free and harmonically trapped Bose gases are given by

$$f_c = -\frac{2\zeta(3)}{\pi} \frac{k_B T}{d^3} \tag{20}$$

and

$$F_c = -\frac{24\zeta(5)\hbar^2}{m^2\omega^2} \frac{k_B T}{d^5}, \tag{21}$$

respectively. It can be found from Eqs. (13) and (20) that the Casimir force per unit area of the free Bose gas in the periodic boundary condition is 8 times the force per unit area in the Dirichlet and Neumann boundary conditions. Moreover, it can be found from Eqs. (16) and (21) that the Casimir force of the harmonically trapped Bose gas in the periodic boundary condition is 32 times that in the Dirichlet and Neumann boundary conditions.

When $T>T_c$, it can be determined from Eq. (17) that $|\varphi_{Cas}(T,\mu)| \le O(e^{-\sqrt{-2\beta\mu}d/\lambda})$, and the Casimir force for the free or harmonically trapped Bose gas falls off exponentially.

Using the above equations, the curves of the reduced scaling functions $\Theta^*$, which vary with the parameter $u$, can be generated, as shown in Fig. 1. Fig. 1 shows that the reduced scaling functions increase monotonically with $u$. When $u=0$, $\Theta_F^* = \Theta_{per,F}^*$ and $\Theta_H^* = \Theta_{per,H}^*$. This result indicates that when $u=0$, the reduced scaling functions are still dependent on the external potential, but they are independent of the Dirichlet, Neumann, and periodic boundary conditions.

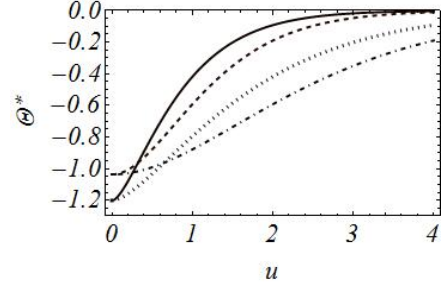

FIG. 1. The curves of the reduced scaling functions $\Theta^*$ varying with $u$, where $\Theta_F^* = 8\pi\Theta_F$ (solid line), $\Theta_H^* = \frac{16}{3}\Theta_H$ (dashed line), $\Theta_{per,F}^* = \pi\Theta_{per,F}$ (dotted line), and $\Theta_{per,H}^* = \frac{1}{6}\Theta_{per,H}$ (dot dashed line).

In conclusion, with the help of the density of states in the generic power-law potential, we have derived the Casimir potential of the ideal Bose gas between two slabs for various boundary conditions. It is found that when $T \le T_c$, the Casimir force is a power-law function that decays with different powers for the different potentials; when $T > T_c$, the Casimir force is an exponential decay function of the separation $d$ of the slabs; when $T \gg T_c$ the Casimir force vanishes. The results obtained are general and can be directly used to discuss the Casimir effect of the ideal Bose gas in the special cases of free and harmonic potentials.

**Acknowledgements**: This work has been supported by the Specialized Research Fund for the Doctoral Program of Higher Education (20100121110024) and the National Natural Science Foundation (No. 11175148), People's Republic of China. Two of the authors (T. Lin and Q. A. Wang) thank the support of the Region des Pays de la Loire in France under the grant number No. 2007-6088 and No. 2009-09333.